\documentclass[aps,prl,reprint,superscriptaddress]{revtex4-1}

\usepackage{graphicx}
\usepackage{color}

\begin{document}

\newcommand{\RL}[1]{\textcolor{red}{#1}}


\title{Inhomogeneous Si-doping of gold-seeded InAs nanowires grown by molecular beam epitaxy}



\author{Chlo\'e Rolland}
\affiliation{Institute of Electronics Microelectronics and Nanotechnology, UMR CNRS 8520, ISEN Department, Avenue Poincar\'e, CS 60069, 59652 Villeneuve d'Ascq Cedex, France}

\author{Philippe Caroff}
\affiliation{Institute of Electronics Microelectronics and Nanotechnology, UMR CNRS 8520, ISEN Department, Avenue Poincar\'e, CS 60069, 59652 Villeneuve d'Ascq Cedex, France}
\affiliation{Department of Electronic Materials Engineering, Research School of Physics and Engineering, The Australian National University, Canberra, ACT 0200, Australia}

\author{Christophe Coinon}
\affiliation{Institute of Electronics Microelectronics and Nanotechnology, UMR CNRS 8520, ISEN Department, Avenue Poincar\'e, CS 60069, 59652 Villeneuve d'Ascq Cedex, France}

\author{Xavier Wallart}
\affiliation{Institute of Electronics Microelectronics and Nanotechnology, UMR CNRS 8520, ISEN Department, Avenue Poincar\'e, CS 60069, 59652 Villeneuve d'Ascq Cedex, France}

\author{Renaud Leturcq}
\email[]{renaud.leturcq@isen.iemn.univ-lille1.fr}
\affiliation{Institute of Electronics Microelectronics and Nanotechnology, UMR CNRS 8520, ISEN Department, Avenue Poincar\'e, CS 60069, 59652 Villeneuve d'Ascq Cedex, France}


\date{\today}

\begin{abstract}
We have investigated \emph{in-situ} Si doping of InAs nanowires grown by molecular beam epitaxy from gold seeds. The effectiveness of $n$-type doping is confirmed by electrical measurements showing an increase of the electron density with the Si flux. We also observe an increase of the electron density along the nanowires from the tip to the base, attributed to the dopant incorporation on the nanowire facets whereas no detectable incorporation occurs through the seed. Furthermore the Si incorporation strongly influences the lateral growth of the nanowires without giving rise to significant tapering, revealing the complex interplay between axial and lateral growth.
\end{abstract}


\maketitle


InAs nanowires (NWs) have been extensively studied for the possibilities to fabricate devices using a low bandgap material, with potential applications for high-performance low-power electronics \cite{Jansson01,Dey01}, high-speed electronics \cite{Blekker01,Egard01}, infra-red opto-electronics \cite{PetterssonH01,WeiW01}, terahertz detection \cite{Vitiello01}, as well as for the strong $g$-factor and large spin-orbit interaction for spintronics \cite{NadjPerge02}. Most of the existing devices are using non-intentionaly-doped NWs, relying on the accumulation layer formed at the surface of InAs for finite conduction, but giving relatively high and uncontrolled resistance due to surface effects \cite{Scheffler01}. Doping of NWs is essential for the realization of electronic and opto-electronic functional devices. In particular, controlled $n$-type doping is an important issue in order to reduce the contact resistance of NW transistors and improve the performances \cite{Dey01} of vertical wrap-gate transistors \cite{Thelander05,Froberg01}, terahertz detectors \cite{Vitiello01} or quantum devices.

$n$-type doping of InAs NWs has been studied using different dopants and techniques, from gold-assisted metal-organic vapor-phase epitaxy (MOVPE) \cite{Thelander04,DoQT01}, chemical beam epitaxy (CBE) \cite{Thelander04,Viti01}, as well as gold-free MOVPE using selective area epitaxy \cite{Wirths01,Ghoneim01}. However, very few studies were aimed at understanding the doping mechanism. Very recently, studies on MOVPE-grown InAs NWs have demonstrated incorporation of Si atoms in the NW core through the gold seed due to the large solubility of Si in gold \cite{ZhangG01}, in contrast to what has been reported using CBE, where mainly radial doping was observed \cite{Thelander04}. $n$-type doping of gold-seeded InAs NWs grown by molecular beam epitaxy (MBE) is missing, in contrast to $p$-type doping \cite{SorensenBS01}.

Here we investigate Si-doping of InAs NWs grown by MBE from gold seed particles using the vapor-liquid-solid mechanism. Si is one of the prefered donors for arsenides and is almost ideal in MBE growth, due to a low vapor pressure (compared for instance with Te), a low diffusion coefficient and a sticking coefficient near unity. 
By using NWs contacted with several electrodes in a back-gated field effect transistor (FET) geometry, we determine the carrier concentration at different positions along the NWs. We observe that the electron density increases linearly from the tip of the NW towards the base, in contrast to a previous study on Si-doped InAs NWs using MOVPE \cite{ZhangG01}. We demonstrate that this inhomogeneous doping is due to incorporation of Si on the side walls, with negligible incorporation through the gold seed. The Si flux also affects the lateral growth of the NWs, but without significant tapering during growth, which raises the question of the incorporation mechanism of the dopants and their influence on the growth of doped NWs using metal-seeded MBE.


\begin{figure}
\includegraphics{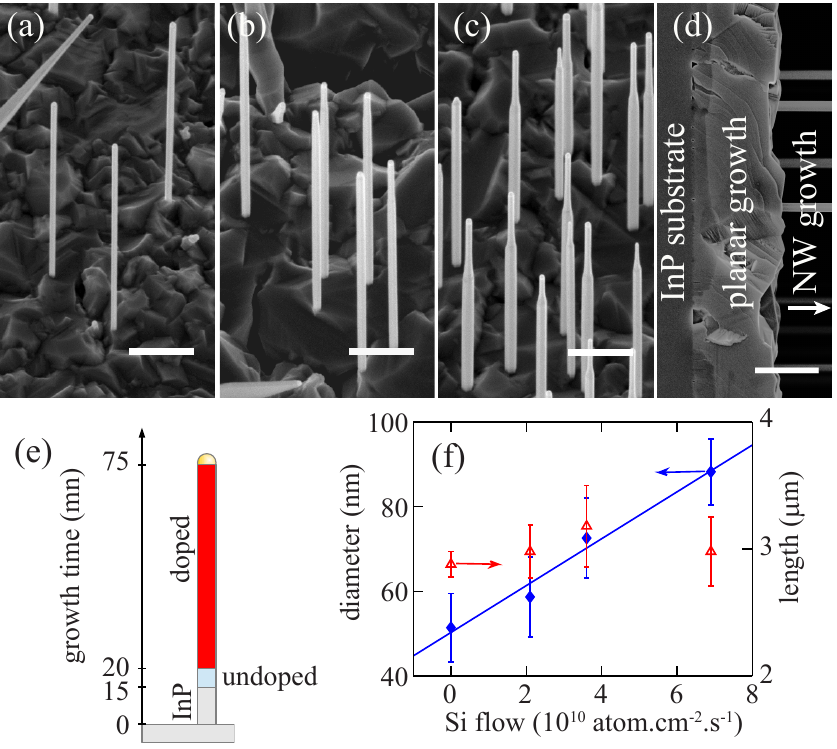}
\caption{ (a-c) SEM pictures with a tilt angle of 30$^\circ$ of as grown NWs for different Si flux: (a) undoped, (b) \emph{low} and (c) \emph{high} doping. (d) Cross-section SEM picture of as grown NWs with medium doping level. The scale bars are 500 nm. (e) Scheme of the growth sequence. (f) Average NW diameter and length as a function of Si flux, extracted from measurements over 10 NWs using SEM pictures similar to (a-c). The diameter is measured about 1 $\mu$m above the surface, and the length measured from the surface to the top of the gold particle. Errors bars are the standard deviations. The solid line is a linear fit of the diameter vs. Si flux. 
\label{fig:nanowiregrowth}}
\end{figure}

All NWs were grown under the same conditions except for nominal Si atomic flux and growth durations, when specified. Growth seeds were pure size selected gold aerosol nanoparticles of 40~nm diameter \cite{Magnusson01} deposited on an InP(111) substrate. As a reference we also used Au droplets self-formed by dewetting at 525~$^\circ$C a nominally 0.2~nm-thick Au film deposited by electron-beam evaporation. Growth was performed at a substrate temperature of $410 \pm 10$~$^\circ$C under V/III ratio below 2 to insure maximal axial growth rate and minimal parasitic radial overgrowth \cite{Tchernycheva02}. As described in Fig.~\ref{fig:nanowiregrowth}(e), growth was initiated by a short InP stem for 15~min., followed by an undoped InAs stem for 5~min.. Then the Si flux is added during a 55~min. long step, before cooling down to room temperature under As$_2$ flux. Some samples have a more complex doped/undoped section with different growth durations, as specified later. We have checked that both short stem segments get buried by concomitant parasitic planar polycrystalline growth ($700 \pm 50$~nm thick for all growth runs, as measured from cross-section images and shown in Fig.~\ref{fig:nanowiregrowth}(d)). Three values of the Si flux have been used, which will be refered as \emph{low} ($2.1 \times 10^{10}$ at.cm$^{-2}$.s$^{-1}$), \emph{medium} ($3.6 \times 10^{10}$ at.cm$^{-2}$.s$^{-1}$) and \emph{high} ($6.9 \times 10^{10}$ at.cm$^{-2}$.s$^{-1}$) doping levels.


The NWs were inspected after growth using a field-emission scanning electron microscope (SEM), see Fig.~\ref{fig:nanowiregrowth}(a-c). Figure~\ref{fig:nanowiregrowth}(f) shows that, as far as morphology is concerned, the Si flux mainly affects the lateral growth leading to a linearly increasing diameter, while the axial growth rate seems to be almost not affected. A lateral overgrowth for Si-doped InAs NWs was already observed using CBE and MOVPE \cite{Thelander04,Wirths01}. We note however that for MOVPE, the larger lateral growth rate is correlated with a smaller axial growth rate \cite{Wirths01}, which is apparently not observed in our case. We do not observe any significant tapering, but the NWs with highest Si flux show a thinner tip of about 0.8 to 1~$\mu$m long, leading to a bottle-like shape. Such a shape has already been observed for MBE-grown NWs as a consequence of finite diffusion length of adatoms on the NW side walls \cite{Plante02,Tchernycheva02,Plante01,Sibirev01}.


\begin{figure}
\includegraphics{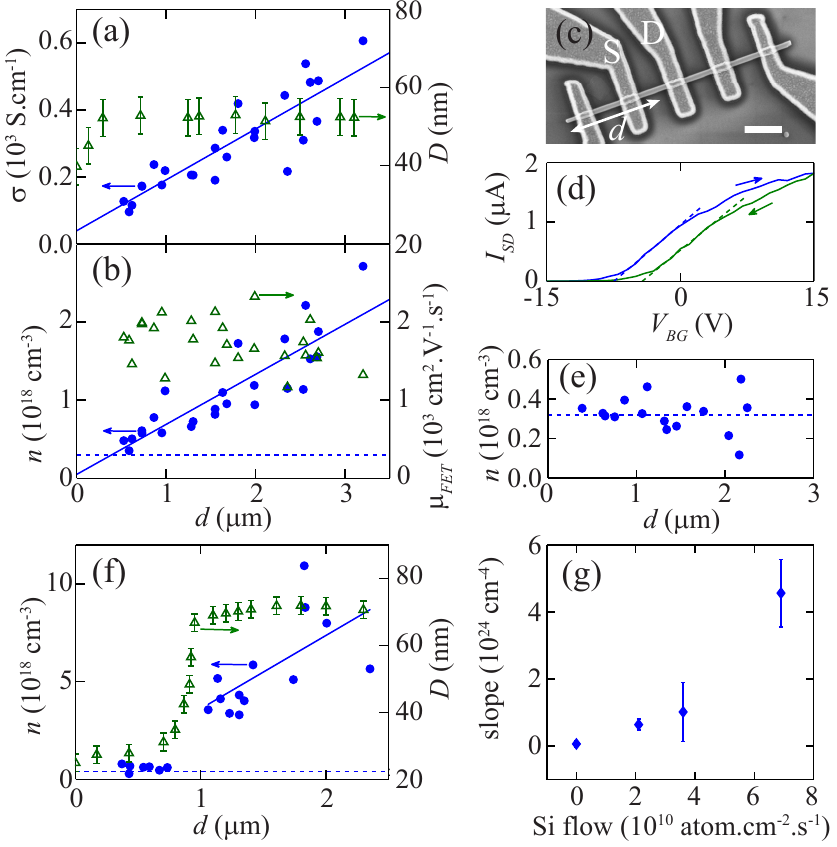}
\caption{(a) Conductivity $\sigma$ (filled dots) and NW diameter $D$ (open triangles), and (b) electron density $n$ (filled dots) and FET mobility $\mu_{FET}$ (open triangles) extracted from $I_{SD}-V_{BG}$ characteristics, as a function of the distance to the gold seed $d$ for a set of NWs grown at \emph{low} doping. The solid line in (a) is a linear fit of $\sigma$ vs. $d$. The error bars result from the resolution of the pictures used for extracting the diameter. (c) SEM picture of a typical NW contacted with five contacts, with the distance $d$ to the gold seed represented. The scale bar is 500 nm. (d) Typical $I_{SD}$ vs. $V_{BG}$ characteristics at a fixed $V_{SD} = 10$~mV taken for increasing and decreasing gate voltage for a NW grown at \emph{low} doping. (e) $n$ vs. $d$ for a set of undoped NWs. (f) $n$ (filled dots) and $D$ (open triangles) vs. $d$ for a set of NWs grown at \emph{high} doping. In (b), (e) and (f) the solid line is a fit of $n$ vs. $d$, and the dashed line represents the average electron density in undoped NWs (see panel (d)). (g) Slope of $n$ vs. $d$ (extracted from linear fit) as a function of Si flux. Error bars are the 95~$\%$ confidence bounds of the linear fit. 
\label{fig:densityvsposition}}
\end{figure}

In order to investigate the effect of Si-doping on the electronic properties, back-gated NW FET devices where fabricated by transferring the NWs to a highly-doped Si wafer to be used as a back gate covered by 225~nm of thermal SiO$_2$. The NWs were located using SEM at low magnification and several contacts with 200 nm width and 500 nm spacing were defined by electron beam lithography (see Fig.~\ref{fig:densityvsposition}(c)), followed by 30~s of O$_2$ plasma ashing, passivation using diluted (NH$_4$)$_2$S$_x$ \cite{Suyatin01}, deposition of 20~nm Ti and 150~nm Au using electron-beam evaporation and lift-off. The multiple contacts geometry allows to provide a local information along the NW \cite{Tutuc03,Imamura01,Dufouleur01}. In order to obtain sufficient statistics and resolution, measurements from 3 to 10 devices containing 3 to 5 contacts are assembled for each growth. Electrical measurements were performed in a vacuum probe station. Using three- and four-contacts measurements we checked that the contact resistance is less than 150~$\Omega$, which is much smaller than typical NW resistance, and the following measurements are thus done in a two-contacts geometry, applying a bias voltage $V_{SD}$ between two consecutive contacts and measuring the current $I_{SD}$.

The conductivity $\sigma$ of each part of a NW is deduced from the low-bias resistance $R$ assuming homogeneous current flow through the section of the NW, $\sigma = 1/R \times L/S$, where $L$ is the length between two contacts and $S=\frac{3\sqrt{3}}{8} D^2$ is the area of the hexagonal section of the NW with $D$ the "tip-to-tip" distance of the hexagon. The conductivity of a set of NWs with \emph{low} doping is shown in Fig.~\ref{fig:densityvsposition}(a) as a function of the distance to the gold seed $d$. We observe a linear increase of the conductivity from the gold particle towards the bottom of the NWs. This behavior has been observed for all doped NWs, and is not observed for undoped NWs. We first checked that this change in conductivity is not related to any change in NW diameter, as shown in Fig.~\ref{fig:densityvsposition}(a). 

In order to identify the origin of the variation of the conductivity vs. $d$, $I_{SD}$ vs. back-gate voltage $V_{BG}$ characteristics were measured in the range $-15$ to $+15$ V at $V_{SD} = 10$ mV (see Fig.~\ref{fig:densityvsposition}(d)). The maximum transconductance $g_m = \max \left( \frac{dI_{SD}}{dV_{BG}} \right)$ and the threshold voltage $V_t$ are deduced by fitting the linear part of the $I_{SD}-V_{BG}$ characteristics (dashed lines in Fig.~\ref{fig:densityvsposition}(d)) and averaged for increasing and decreasing $V_{BG}$ to account for an hysteresis due presumably to charging of traps in the oxide or at the NW-oxide interface \cite{OngHG01}. The electron density is calculated following $n = C_{ox} V_t/(|e| S)$, with $e$ the electron charge and $C_{ox}$ the oxide capacitance. We note here that $n$ is thus an average value over the section of the NW. For the capacitance we use the formula $C_{ox} = \alpha \left( 2 \pi \varepsilon_0 \varepsilon_{r}^{\star} \right) / \cosh^{-1} \left( (t_{ox} + \frac{\sqrt{3}}{2}r^{\star})/r^{\star} \right)$, with $\varepsilon_{r}^{\star} = 2.53$ and $r^{\star} = 0.42 D$ for an hexagonal-shaped NW lying on SiO$_2$ and half-surrounded by vacuum \cite{Wunnicke01}, and the coefficient $\alpha = 0.60$ takes into account screening by the source-drain contacts separated by $L = 500$~nm \cite{Note1}. The FET mobility is deduced from $g_m$ using $\mu_{FET} = L g_m / (C_{ox} V_{SD})$.

As shown in Fig.~\ref{fig:densityvsposition}(b), the main origin of the variation of the conductivity as a function of the distance to the tip is the variation of the electron density, while the mobility remains almost unchanged. This change in electron density is not observed for undoped NWs, as shown in Fig.~\ref{fig:densityvsposition}(e), where an average electron density of $3 \times 10^{17}$ cm$^{-3}$ is attributed to the surface pinning of the Fermi level in the conduction band. 
Similar linearly increasing electron density as a function of $d$ is observed for all Si fluxes tested in this study, as show e.g. in Fig.~\ref{fig:densityvsposition}(f) for the highest one. An interesting behavior observed on these NWs with a bottle-like morphology is that only the thicker part of the NW is doped while the thinner top part remains at the doping level of undoped NWs. 
Figure~\ref{fig:densityvsposition}(g) shows that the slopes of $n$ vs. $d$ increases as a function of the Si flux, which is a proof of the efficient \emph{in-situ} Si-doping of InAs NWs by MBE.

\begin{figure}
\includegraphics{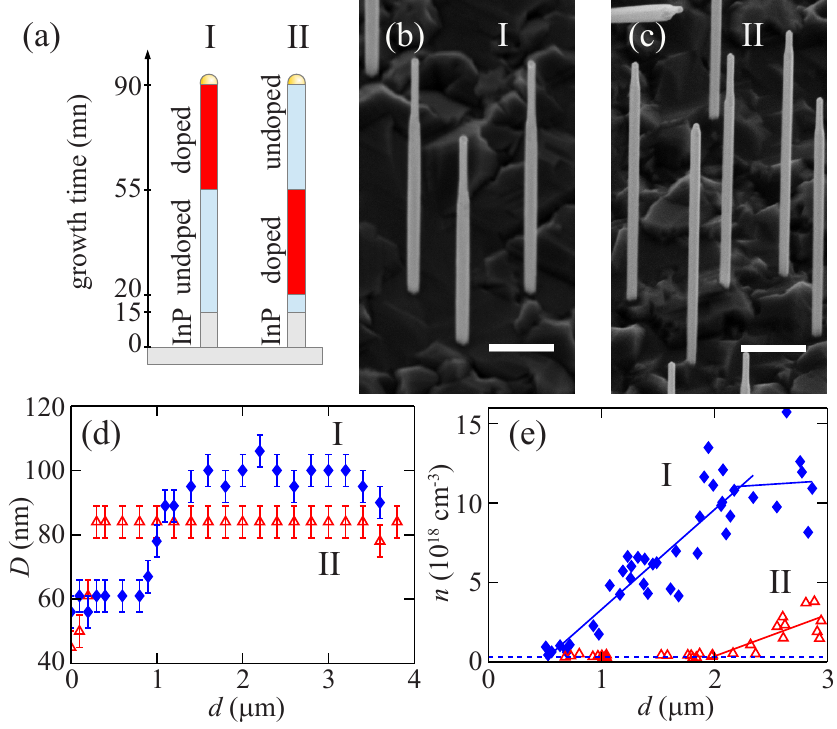}
\caption{ (a) Scheme of the growth sequence for the NWs of types I and II: type I with bottom grown undoped and the top grown with Si flux, and type II with bottom grown with Si flux and top undoped. (b-c) Scanning electron microscope pictures with a tilt angle of 30~$^\circ$ of as grown nanowires of (b) type I and (c) type II for \emph{high} doping. The scale bars are 500 nm long. (d) Nanowire diameter as a function of position compared to the tip for two typical NWs of type I (filled diamonds) and type II (open triangels). (e) Electron density $n$ extracted from the threshold voltage as a function of the distance $d$ to the gold seed for NWs of type I (filled diamonds) and type II (open triangles). The solid lines are linear fits of parts of each data set, and the dashed line represents the average electron density in undoped NWs (see Fig.~\ref{fig:densityvsposition}(d)).
\label{fig:twoparts}}
\end{figure}

In order to investigate the origin of the inhomogeneous carrier density, we have grown two types of NWs with two sections. In type I, first an undoped section is grown followed by a doped section. In type II, we start first with a doped section, followed by an undoped one (see Fig.~\ref{fig:twoparts}(a)). For the doped section, we have used both \emph{medium} and \emph{high} doping, showing similar trends. In the following we present the results obtained for \emph{high} doping. First we note in Fig.~\ref{fig:twoparts}(b-d) that both growths produce NWs with negligible tapering, except for a thin tip (bottle-like shape) as already observed for high doping.

The measurements of electron density presented in Fig.~\ref{fig:twoparts}(e) show that, when the undoped section is grown first (type I), we find a high electron density in the nominally undoped bottom part (more than 30 times larger than undoped NWs). Furthermore the electron density is constant in the undoped section, and decreases in the doped section, as emphasized by the two linear fits in Fig.~\ref{fig:twoparts}(e). For type II NWs, after the growth of a doped part, stopping the Si flux removes any trace of doping (see Fig.~\ref{fig:twoparts}(e)), showing that there is no diffusion of dopants along the NW, nor transient regime due to the seed \cite{Connell01}. Surprisingly we observe that the NW diameter in the undoped part is larger than the one obtained for undoped NWs, and we see no indication of diameter change between both parts.


We first discuss the incorporation of dopants in our system. For vapor-liquid-solid growth, dopants can be incorporated either from the metal seed \cite{Schwalbach01,Yang01}, from the side walls of the NW, or a combination of both. The high electron density measured in the undoped part of type I NWs in Fig.~\ref{fig:twoparts}(e) shows that finite amount of Si is incorporated from the side walls, as observed in Si-doped InAs NWs grown by CBE \cite{Thelander04}. Additionaly, the absence of doping measured in the thinner neck region of bottle-shaped NWs obtained for \emph{high} doping (see Fig.~\ref{fig:densityvsposition}(f)) shows that negligible amount of Si is incorporated from the gold seed. This is confirmed in measurements of type I NWs, where an even partial incorporation from the seed would lead to an upward step in electron density when passing from the undoped to the doped part, while we observe a continuous decrease.

Another important result of our investigation concerns the morphology of the doped NWs. It is clear that undoped NWs have negligible lateral overgrowth, while doped NWs have a larger final diameter, except for a final bottle-shaped top part close to the seed particle, thus revealing that Si doping induces signifant lateral growth but no tapering. We suggest that in the case of optimised growth parameter and absence of Si flux, nucleation on the wurtzite sidewalls is largely cancelled and only axial growth exists. When Si is incorporated at the sidewalls, it reduces the adatom diffusion length, thus favoring a few nucleation events on the sidewalls. However, the homogeneous diameter observed for all doped NW samples proves that step-flow is still the dominant radial growth mechanism under these conditions \cite{Plante01,Sibirev01}. Sample II (a doped segment followed by an undoped segment), which shows an homogeneous large diameter up to the top, even though the doped segment is only half of the length, confirms that once initiated, taper-free step-flow overgrowth is maintained while Si dopant flux is suppressed, resulting in NWs with highly homogeneous diameters. It also suggest that NW shape could evolve continuously, thus hiding originally tapered shapes.

Finaly we discuss the possible origin of the linear increase of the carrier density as a function of the distance to the tip. Such a dependence is usually expected from incorporation of dopants from the side walls since the bottom of the NW is longer exposed to the dopant flux compared to the top, as opposed to doping from the seed which usually leads to increasing doping towards the tip \cite{Gutsche01,Dufouleur01,ZhangG01}. A first possibility could be the growth of a doped shell with a decreasing thickness from the bottom to the top \cite{Tutuc03,Dufouleur01}. This scenario cannot apply directly to our case, because we do not observe significant tapering \cite{Note2}, although concomitant step-flow growth could alter the final NW shape. Another possible mechanism is the incorporation of dopants and diffusion towards the core as a source of inhomogeneous doping without significant tapering \cite{AllenJE01,Koren02,Casadei01}.


In conclusion we report on \emph{in-situ} Si-doping of InAs nanowires using gas source MBE. We demonstrate the inhomogeneous doping mechanism, with an electron density increasing from the nanowire tip towards its base, while no significant tapering is observed. We attribute this result to incorporation of Si from the NW side walls, with negligible incorporation through the seed. The exact incorporation mechanism leading to lateral doping with no tapering remains unclear, but could be related to the step-flow growth mechanism explaining radial growth with no tapering. Future studies will aim at compensating for the axial carrier concentration gradient, and modifying the incorporation pathways between particle and sidewalls via growth parameter tuning.

This work was supported by the ANR through the Project No. ANR-11-JS04-002-01, and the Ministry of Higher Education and Research, Nord-Pas de Calais Regional Council and FEDER through the “Contrat de Projets Etat Region (CPER) 2007-2013.” We thank C. Boyaval for support on cross-section SEM imaging, and K. A. Dick for the aerosol-generated gold seed particles. P.C. is the recipient of an Australian Research Council Future Fellowship (project number FT120100498).




%

\end{document}